\documentclass[final,3p,times,twocolumn,authoryear]{elsarticle}

\usepackage{amssymb}
\usepackage{url}
\usepackage{hyperref}
\usepackage{xcolor}
\hypersetup{
    colorlinks=true,
    }
\usepackage{geometry}
 \usepackage{changepage}

\usepackage{setspace}
  
\usepackage{amsthm}
\usepackage{enumerate}
\usepackage{amsmath}
\makeatletter
\g@addto@macro\normalsize{
  \setlength\abovedisplayskip{30pt}
  \setlength\belowdisplayskip{30pt}
  \setlength\abovedisplayshortskip{30pt}
  \setlength\belowdisplayshortskip{30pt}
}
\makeatother
\usepackage{float}
\usepackage{natbib}
\usepackage{cite}
\usepackage{textcomp}
\usepackage[export]{adjustbox}

\begin{document}
\begin{frontmatter}

\title{Radio Observation of Venus at Meter Wavelengths using the GMRT}

\author[nithin]{Nithin Mohan}
\author[subhashish]{Subhashis Roy}
\author[subhashish]{Govind Swarup}
\author[subhashish]{Divya Oberoi}
\author[subhashish]{Niruj Mohan Ramanujam}
\author[nithin]{Suresh Raju C.\corref{cor1}}
\cortext[cor1]{Corresponding author}
\ead{c$\_$sureshraju@vssc.gov.in}
\author[nithin]{Anil Bhardwaj}
\address[nithin]{Space Physics Laboratory, Vikram Sarabhai Space Center, 
Thiruvananthapuram-695022, India.}
\address[subhashish]{National Center for Radio Astrophysics, Tata Institute of 
Fundamental Research, Pune-411007, India.}

\begin{abstract}

The Venusian surface has been studied by measuring radar reflections and thermal radio 
emission over a wide spectral region of several centimeters to meter 
wavelengths from the Earth-based as well as orbiter platforms. The radiometric 
observations, in the decimeter (dcm) wavelength regime showed a decreasing
trend in the observed brightness temperature ($\mathit{T}_\textup{b}$) with increasing wavelength.
The thermal emission models available at present have not been able to explain the radiometric
observations at longer wavelength (dcm) to a satisfactory level.
This paper reports the first interferometric imaging observations of Venus below 620 MHz.
They were carried out at 606, 332.9 and 239.9\,MHz using the Giant Meterwave Radio Telescope (GMRT). 
The $\mathit{T}_\textup{b}$ values derived at the respective frequencies
are 526\,K, 409\,K and $<$\,426\,K, with errors of $\sim$7\% which are
generally consistent with the reported $\mathit{T}_\textup{b}$ values at 608\,MHz and 430\,MHz
by previous investigators, but are much lower than those derived from high-frequency
observations at 1.38-22.46\,GHz using the VLA.
\end{abstract}

\begin{keyword}
Venus, surface\sep Radio Observations\sep Radiative transfer
\end{keyword}

\end{frontmatter}
\section{INTRODUCTION}

Venus being the nearest planet, its dense atmosphere and the surface have been
the subject of many studies, including the ones from orbiting spacecraft, landers and
from Earth-based observations in the last six decades:

\begin{enumerate}[(i)]

\item orbiting spacecraft,
Mariner \citep{Fjeldbo1971}; Magellan \citep{Pettengill1991};
Venus Express \citep{Markiewicz2007}
\item Venera and Vegas Landers
(\citeauthor{Vinogradov1976},\,\citeyear{Vinogradov1976}; 
\citeauthor{Florenskii1982},\,\citeyear{Florenskii1982};
\citeauthor{Basilevsky1986},\,\citeyear{Basilevsky1986}), and 
\item radar and radio observations made from the Earth 
(\citeauthor{Campbell1989},\,\citeyear{Campbell1989}; 
\citeauthor{Condon1973},\,\citeyear{Condon1973}; 
\citeauthor{Muhleman1973},\,\citeyear{Muhleman1973}; 
\citeauthor{Butler2001},\,\citeyear{Butler2001}).

\end{enumerate}

\par  The atmosphere of Venus is comprised of $\sim$\,96\% CO$_2$, $\sim$\,4\% 
of N$_2$ and trace amounts of gases like H$_2$O, SO$_2$, CO and H$_2$SO$_4$ 
(\citeauthor{Seiff1980}\,\citeyear{Seiff1980};
\citeauthor{Butler2001}\,\citeyear{Butler2001}). The thick atmosphere generates 
a pressure of $\sim$\,90 bars at the surface. 
Since CO$_2$ is a very efficient greenhouse gas, the surface is extremely hot, 
$\sim$735\,K; and for the same reason, it does not have a significant diurnal as well 
as equator to pole variation of temperature. Many studies have reported possible 
characteristics of the interior of Venus and its tectonic nature, as well as the
possible areas of the Venus surface that expel its internal heat 
(\citeauthor{Sinclair1970},\,\citeyear{Sinclair1970};
\citeauthor{Marov1978},\,\citeyear{Marov1978}; 
\citeauthor{Seiff1980},\,\citeyear{Seiff1980}; 
\citeauthor{Phillips1983},\,\citeyear{Phillips1983}). It is postulated that 
heat generation from its core and by radioactive elements is similar to that for the 
Earth or Mars. \citet{Surkov1983} has summarized studies of Venus rocks by Venera 8, 9 
and 10 that landed on its surface during 1972-75. The mean contents were found to be 
close to that of the 
basalts and granites of the Earth's crust having a density of 2.8\,$\pm$\,0.1 
g\,cm$^{-3}$.

\par The Venusian surface has been explored since 1961 with the first radar 
observation of Venus from Earth  carried out at NASA's Goldstone Observatory 
(\citeauthor{Carpenter1964},\,\citeyear{Carpenter1964}).  
The successive radar observations revealed important information about Venus,
such as its rotation is retrograde and the rotation period is 243.1 days,
its axis of rotation is almost perpendicular to its orbital plane, and the planetary  
radius is $\sim$\,6,052\,km (\citeauthor{Goldstein1963},\,\citeyear{Goldstein1963}).

\par Besides the ground-based radar observations, Venus surface has been mapped using 
spacecraft-based radars on the Pioneer 10 
(\citeauthor{Pettengill1980},\,\citeyear{Pettengill1980};
\citeauthor{Muhleman1979},\,\citeyear{Muhleman1979}; 
\citeauthor{Ford1983},\,\citeyear{Ford1983}); and the Magellan probe 
(\citeauthor{Pettengill1991},\,\citeyear{Pettengill1991}; 
\citeauthor{Campbell1994},\,\citeyear{Campbell1994}). 
All these studies were limited to single frequency and in horizontal (H) polarization.  
Studies of radar echoes at longer wavelengths ranging from 15\,cm to 7.84 m 
showed global mean surface reflectivity values of $\sim$\,0.15. The low 
reflectivity of 
$\sim$\,0.02 at 3.8\,cm was attributed to the attenuation of radar echoes by the
atmosphere at these higher frequencies
\citep{Pettengill1980}. These radar measurements also indicated the presence of a 
thin layer (of $\sim$\,centimeter thickness) of porous powdered soil or dust 
\citep{Pettengill1988}. The Magellan radar (SAR) data have been used extensively 
to characterize the Venusian surface by 
studying geomorphology and variation in the dielectric properties of the
high and lowland regions. The dielectric permittivity at lowlands are 
$\sim$\,5 showing the presence
of dry basaltic or granite minerals, but high value of dielectric permittivity 
($>$\,50) at highlands indicates the presence of highly conducting mineral 
deposits or the presence of less absorbing materials that can return most 
of the incident radar signals back at these locations. 
These radar investigations showed that about 15\% of the impinging radiation is 
reflected, indicating that the dielectric permittivity of Venus is in the range 
of $\sim$\,4.15 to $\sim$\,4.5 \citep{Pettengill1992}.

\par Several investigations have been made to derive properties of the thermal 
radio emission of Venus using Earth-based and space-borne radio telescopes operating 
at centimeter and decimeter wavelengths 
(\citeauthor{Kuzmin1983},\,\citeyear{Kuzmin1983};
\citeauthor{Condon1973},\,\citeyear{Condon1973}; 
\citeauthor{Muhleman1979},\,\citeyear{Muhleman1979};
\citeauthor{Pettengill1992},\,\citeyear{Pettengill1992}; 
\citeauthor{Butler2001},\,\citeyear{Butler2001}).
The  passive mode operation of the Magellan radar enabled the measurement of 
radio emissivity at 
12.6\,cm wavelength in horizontal polarization for more than 91\% of the Venus 
surface during the first 8 months of
its operation \citep{Pettengill1992}. The global mean value of emissivity observed using 
horizontal linear polarization is 0.845, a value that corresponds to a 
dielectric permittivity
of between 4.0 and 4.5, depending on the surface roughness. These values are consistent with
the permittivity values of dry basaltic minerals that compose the bulk of the Venus 
surface. The above emissivity value is also in good agreement with that derived from the 
radar reflectivity.

\par \citet{Kuzmin1983} summarized early radiometric measurements of the 
brightness temperatures ($\mathit{T}_\textup{b}$)
of Venus made using Earth-based radio telescopes. \citet{Butler2001}  
measured Venusian 
$\mathit{T}_\textup{b}$ at different frequencies by using the VLA observations at
22.46, 14.94, 8.44, 4.86 and 
1.385\,GHz with errors of $\sim$\,2 to 5\%. 
At shorter wavelengths ($<$\,5 cm or frequencies above 6\,GHz) radiation arises 
primarily from the dense 
atmosphere of Venus but at longer wavelengths (decimeter and meter wavelengths) 
thermal radiation
is increasingly dominated by the surface and subsurface emission 
\citep{Warnock1972}.
\citet{Butler2001} proposed a detailed model, to explain the VLA measured values 
of $\mathit{T}_\textup{b}$ considering absorption at microwave frequencies by the 
atmosphere based on
the vertical profiles of SO$_2$ and H$_2$SO$_4$ derived from the Pioneer-Venus 
(PV) probes and that
inferred from measurements of the Mariner V (MV) spacecraft 
\citep{Muhleman1979}. At longer wavelengths, the
contribution of radiation from the surface and subsurface of Venus is also 
considered. However, their 
model predicts $\mathit{T}_\textup{b}$ values much higher than those measured
at 1.385\,GHz and at $<$\,1\,GHz by others.
\citet{Muhleman1973}, and \citet{Condon1973} measured $\mathit{T}_\textup{b}$ 
values at frequencies $<$\,1\,GHz and found them to be in the range 500-550\,K. 
These $\mathit{T}_\textup{b}$ values are significantly lower than those measured at higher 
frequencies. 
The radiometric measurements made during 1972\,-\,73 used Wyllie's flux density 
scale for calibrations and doubts were raised in the literature  about the  
scale used those measurements at frequencies $<$\,1\,GHz \citep{Condon1973}.
However, subsequently \citet{Baars1977} noted that \textquotedblleft Wyllie's flux density scale
is only 3\% above our CasA scale\textquotedblright.
It is to be noted that \citet{Baars1977} flux density scale is widely used today by radio
astronomers across the world for flux density calibration.

\par The depth of penetration of micro/radiowaves into the Venusian regolith 
depends on the dielectric properties of the same. The lander based in-situ 
measurements, Earth-based radar/radiometric measurements as well as the 
orbiter measurements concluded that the Venusian surface is dry and
has low dielectric constant values of $\sim$\,4.5. With this consideration the
observation at meter wavelengths are suitable for the study of deeper 
subsurface characteristics since the penetration depth, $\delta$ (in meters),
of the radiation, is related to the wavelength by the equation 
\begin{equation}
 \delta  = \frac{\lambda _0 \sqrt{\varepsilon'}}{2\pi \varepsilon''} 
\end{equation}
where $\varepsilon'$ is the real part of the and $\varepsilon''$ is the 
imaginary part of the dielectric and $\lambda _0$ is the wavelength in free space.
It has been found that there is a significant decrease in $\mathit{T}_\textup{b}$ beyond a 
wavelength of $\sim$15\,cm. There were a large number of successful observations carried 
out between several millimeters (mm) to centimeter (cm) wavelength. But no 
observations were reported beyond 70\,cm ($\lesssim$\,400 MHz) due to an
increased system and background noise, solar interference and weak planetary emission
\citep{Condon1973}. 

\par Flux density measurements based on interferometric imaging do not suffer from
base-level variations, solar interference and local radio-frequency interference 
which often afflict single dish observations \citep{Condon1973}, the 
latter two can also afflict the non-imaging interferometric observations 
\citep{Muhleman1973}. 
Here we report on interferometric observations carried out at 3 different
wavelengths 50\,cm, 90\,cm and 123\,cm (or 606\,MHz, 332.9\,MHz and 239.9\,MHz, 
respectively) using Giant Meterwave Radio Telescope (GMRT). The $\mathit{T}_\textup{b}$ of
Venus measured from images made from these observations can serve as inputs
for developing an improved thermal emission model that can account for the 
increasing subsurface thermal emission at longer wavelengths. These are the
first reported interferometric imaging flux density and $\mathit{T}_\textup{b}$
of Venus at frequencies lower than 620\,MHz. 

\par This paper presents the results obtained from analyzing the archival data
of Venus (project code 05BBA01) collected during the observations made
in March 2004 using the GMRT at 239.9\,MHz, 332.9\,MHz and 606\,MHz, as
discussed in Sections 2 and data processing and reduction are presented in 
Section 3. The results are presented in Section 4, followed by discussion and 
conclusions in Sections 5 and 6, respectively.

\section{OBSERVATIONS} 
Observations of Venus were carried out on six days between March 20 and March 27, 2004
using the GMRT at three frequencies centered close to  239.9\,MHz, 332.9\,MHz and 606\,MHz.
The 239.9\,MHz and 606\,MHz the observations were conducted simultaneously using the dual
frequency co-axial GMRT feeds.
The design of the GMRT is described in \citet{Swarup1991}.
Briefly, GMRT consists of 30 fully steerable parabolic dishes each of 45 m in diameter,
14 of which are located in a central array of $\sim$1\,km $\times$ 1\,km extent and the other 
16 in a Y-shaped array of extent 25 km.
Two of the six days of observations had short observing runs, only the observations from the 
other four days are presented here.
An integration time of 16.8\,second and a spectral resolution of 125\ kHz was used for these observations.
Table 1 summarizes the observation details. 

Both 3C147 and 3C48 were used as primary flux calibrators at 606.0, 606.1 and 239.9 MHz and
only 3C48 was used for this purpose at 332.9 MHz.
The primary flux calibrators were observed at the start and end of each observing session.
The flux density of the flux calibrators was obtained from the task SETJY in the 
Astronomical Image Processing System (AIPS) and Common Astronomy Software Applications (CASA), 
which use Scaife and Heald flux density scale \citep{Scaife2012} below 500\,MHz 
and Perley and Butler scale \citep{Perley2013} above 500\,MHz.  
These scales are in close agreement with \citet{Baars1977} scale at frequencies $\sim$ $>$\,300\,MHz. 
The compact radio source 0318+164 was used as both the phase and bandpass calibrator,
it was observed for $\sim$5 mins every 30 minutes.

Rather than tracking Venus, whose right ascension (RA) and declination (Dec) change with time,
the antennas tracked the RA, Dec corresponding to the coordinates of Venus at the middle of
observing period for that particular day. 
The ephemeris details of the Venus during the observations are tabulated in Table 2. 
The diameter of Venus varied from 21.19 arcsec to 22.97 arcsec over the 6 days of observations.

\begin{table*}[!t]
  
  \caption{GMRT observations of Venus 2004} 
  \centering\small
  \begin{tabular}{ccccc}
\\
\hline
 
  Central   & Date  & Bandwidth & No. of   & Observing    \\
  Frequency &	    &    (MHz)  & working  & time on	  \\
  (MHz)	    &	    &   	& antennas & Venus (mins) \\
  \hline 
  
    606.1 & March 26 & 16 	& 26 	   & 311	\\
    606.0 & March 27 & 16 	& 26 	   & 334	\\
    332.9 & March 19 & 16 	& 28 	   & 210	\\
    332.9 & March 20 & 16 	& 27 	   & 240	\\
    239.9 & March 26 & 8  	& 26 	   & 311	\\

  \hline	      
  \end{tabular}	      
\end{table*}

\begin{table*}[!t]
\caption{Ephemeris data for GMRT observations of Venus 2004}
\begin{center}
\centering\small
\begin{tabular}{cccccc}
  \hline
   Central 	 &  	Date	&  Venus Angular 	 & Right Ascension 	& Declination		&	\% illumination	\\
   Frequency	 & 		&  Diameter\,($\theta$)  & (mid-observation)	& (mid-observation)	&			\\
   (MHz)	 & 		&  (arcsec)	 	 & (h:min:s)	
				&  (deg.arcmin.arcsec)	 &									\\
 \hline
  606.1		&	March 26	&  22.72	 &	3:14:48	&	+21.05.00	&	52.9			\\
  606.0		&	March 27	&  22.97	 &	3:18:59	&	+21.25.33	&	52.3			\\
  332.9		&	March 20	&  21.40	 &	2:50:35	&	+18.57.51	&	55.9			\\
  332.9		&	March 19	&  21.19	 &	2:46:30	&	+18.35.06	&	56.4			\\
  239.9		&	March 26	&  22.72	 &	3:14:48	&	+21.05.00	&	52.9			\\
		
  \hline
  \end{tabular}

  \end{center}
\end{table*}

\section{DATA REDUCTION}

\subsection{Analysis Methodology}
To build confidence in our analyses and results, we use both CASA and AIPS and
somewhat different analysis procedures.
The observations made at 606.1 and 239.9\,MHz were analyzed using both AIPS and CASA, 
those at 332.9\,MHz were 
analyzed only using AIPS and those at 606.0\,MHz were analyzed only using CASA.
Data editing was performed to remove records highly deviant in amplitude, arising from 
man-made radio interference, in both time and frequency domains. 
The flagging of data analyzed in CASA was carried out using a combination of automated 
flagging outside CASA using FLAGging and CALibration (FLAGCAL), 
a software pipeline developed to automate the flagging and calibration of GMRT
(\citeauthor{Prasad2012},\,\citeyear{Prasad2012}; 
\citeauthor{Chengalur2013},\,\citeyear{Chengalur2013}), and manual flagging in CASA. 
The data analyzed in AIPS were first manually flagged for dead antennas followed by the 
automated flagging task RFLAG. First, we describe the analysis procedure followed in CASA. 
As mentioned earlier, we tracked the mean RA, Dec of Venus for every observing day. 
While this allowed Venus to move appreciably with respect to the phase center and in the antenna
beam over the course of the observations, 
tracking at the sidereal rate enables us to do self-calibration to correct for the phase changes
introduced by the GMRT electronics and the ionosphere during the observations. 
The background celestial radio sources seen in the field of view were used for self-calibration.
Next, CASA task UVSUB was used to subtract contribution of the background sources from the
self-calibrated visibility data using the statistically significant deconvolved 
(CLEAN) components in the model.
An often used technique for imaging a source with non-sidereal motion (Venus) is
to make multiple individual images,
each of them over a duration short enough that the angular displacement of the
source in that period is significantly smaller than the resolution of the imaging instrument.
This method was followed for the 606.0, 606.1\,MHz and 239.9\,MHz observations. 
Synthesized beams at 606.0 and 606.1\,MHz were about 6 arcsec and about 15 arcsec at 239.9\,MHz. 
The angular velocity of Venus in RA and Dec was about 2.65 arcsec/minute and individual
images were made every one minute. 
After primary beam correction, all the one-minute maps were aligned using the position
of Venus available from the NASA JPL Horizons ephemeris
(\url{http://ssd.jpl.nasa.gov/horizons.cgi}) and co-added.
Finally, the co-added map of Venus was deconvolved using a point-spread-function
corresponding to unflagged data for the entire duration of the observation.

The analysis strategy used in AIPS is common with that followed in CASA till self-calibration
and subsequent removal of the contribution of background sidereal sources using UVSUB.
The motion of Venus across the beam is accounted for by using a different strategy.
An artificial strong point-source (10 Jy) was added to the last spectral channel of the dataset. 
The position of this artificial source was shifted every minute in a direction equal and opposite
to that of movement of Venus in the sky plane. 
Then, using a script, phase-only self-calibration was carried out only on the last spectral 
channel of this dataset using this artificial point source and a solution interval of one minute. 
The antenna solutions determined arise entirely due to the motion of the artificial
source and were applied to all the frequency channels of the UV dataset. 
This changed the visibility phases of the UVSUB-ed data precisely by the amount needed to
compensate for the motion of Venus. 
The last spectral channel, where the artificial source was introduced, was not used for imaging of Venus.
Both of these analyses at 606.1\,MHz on 26 March, 2004, using independent 
software suits and differing procedures, yielded flux densities which differ by $\le$\,5\% with similar images.

\subsection{Flux calibration uncertainties}

Given the challenging nature of these observations, particular attention was paid to
estimating the uncertainties in flux calibration.
In general, the errors in flux density of Venus as measured by an interferometer comprising 
physically large antennas like the GMRT arise from the following main reasons:
(i) instrumental \textquoteleft gain\textquoteright \,variations due to the change in pointing
directions between the primary and secondary calibrator, and also between the 
secondary calibrator and Venus; (ii) gain variations over time; and (iii)
uncertainty in absolute flux density scales used to estimate the flux density of the primary calibrator.
Each of these concerns is discussed briefly below. To quantify the net observed variation
of the antenna gains during these observations, 
we measured the gains of the antennas from uncalibrated data towards the secondary calibrator 
observed on 19 March, 2004 at 332\,MHz. Over the 5 hours of observations of the 
secondary calibrator, a gain variation of $\sim$10\% was found.
Typically for large dishes, the gain changes with elevation angle and these gain 
variations are correlated across different antennas. 
For the purpose of estimating uncertainties in flux density measurement, we assume
these variations to be 100\% correlated. 
Given that the flux density is directly proportional to the square of the gain, 
the resulting uncertainty in the flux 
density estimates cannot be larger than $\sim$20\%.
However, observing the primary and secondary calibrators at similar elevations and frequent 
observations of the secondary calibrator reduce this uncertainty substantially
as discussed later in this section.

We make the conservative assumption of the gain change being linear with elevation angle.
During the observations, the elevation of the secondary calibrator changed 
from $\sim$85$^{\circ}$ to $\sim$23$^{\circ}$ (a change of $\sim$62$^{\circ}$). 
The primary calibrator was observed at the start of observation at an elevation angle of 67$^{\circ}$.
Attributing the entire observed gain variation of $\sim$10\% over the elevation
range covered to elevation angle dependence of GMRT dishes, the difference between 
the elevation angles of the two calibrators implies a gain change of $\sim$2.5\%. 
As the angular distance between Venus and the secondary calibrator was much smaller ($\sim$8$^{\circ}$),
a similar argument leads to an expected elevation angle dependent gain
variation between these two sources to be $\sim$1\%.

In addition, the instrumental gains might also drift over time. 
The gains vary slowly and smoothly in time and we assume this variation to be
proportional to the difference in time over time scales of interest here. 
In order to track these gain variations, the observations of primary and secondary
calibrators were done within 30 minutes of each other and the secondary calibrator 
was observed every 30 minutes. Assuming all of the 10\% of the gain variation to
come from such gain drifts in time, the estimated change in gain due to the time difference
while observing primary and secondary calibrator, and secondary calibrator and Venus is $<$1\%.

The absolute flux density scale is now believed to be accurate to better than 3\% \citep{Scaife2012}.
Since all the above-mentioned errors add in quadrature, we finally get an error in
measured flux density of Venus to be $\sim$7\%. We note that in practice the elevation
dependent gain variation is shallower than linear dependence assumed here. It varies 
much more slowly near the zenith, where the
absolute gain from the primary was used to calibrate the flux density of the secondary 
calibrator. In addition, the gain variations in time of each antenna are independent 
and hence expected to contribute randomly to the uncertainty in the measured flux
density. Both of the above-mentioned factors will reduce the uncertainty in flux
density estimate when compared to the estimate presented.

\subsection{Uncertainties in Galactic background temperature}
The full sky map at 408\,MHz made by \citet{Haslam1982} was used to estimate the
Galactic background temperature, $\mathit{T}_\textup{gal}$, towards the direction of Venus. 
The 408\,MHz $\mathit{T}_\textup{gal}$ towards Venus was 30\,K during our observations. 
Considering the uncertainty in the zero level and absolute calibration, the 
uncertainty on \citet{Haslam1982} measurement is $\sim$4\,K. 
The spectral index of $\mathit{T}_\textup{gal}$ near the location of Venus
is measured to be -2.6$\pm$0.15 \citep{Reich1988}.  
This leads to a $\mathit{T}_\textup{gal}$ of 10$\pm$1.5, 52$\pm$7 and 122$\pm$22\,K at 
606, 332.9 and 239.9\,MHz, respectively.

The position of Venus itself in the sky changed by $\sim$5$-$10$'$ during observations 
and the $\mathit{T}_\textup{gal}$ can vary as a function of the Galactic latitude and longitude (l, b). 
The \citet{Haslam1982} map has a resolution of 0.85$^{\circ}$ and any variation 
of $\mathit{T}_\textup{gal}$ at smaller angular scales is averaged out in this map.  
However, $\mathit{T}_\textup{gal}$ variations at angular scales of 1$-$30$'$ are
easily picked up by GMRT 330 MHz band observations. 
To estimate this variation, we subtracted out the background extragalactic sources 
at high resolution and then made a low-resolution map of the region at 332 MHz.  
The resultant map had a resolution of 159$^{''} \times 135^{''}$. 
The \textit{rms} of the map was $\sim$3~mJy/beam, which corresponds to a 
fluctuation of $<$\,2\,K in $\mathit{T}_\textup{gal}$ at 330 MHz at $\sim$3$-$30$'$ scales. 
No structures at scales $>5'$ were seen with a significance $>2\sigma$. 
Therefore, the uncertainty due to angular variations in $\mathit{T}_\textup{gal}$ scaled 
to the above frequencies is low in comparison to the base level uncertainty in \citet{Haslam1982} map.

\subsection{Contamination from background sources}
Most background sources tend to have spectral indices which will make them 
brighter at lower frequencies, where we find Venus to be weaker. 
So we have carefully examined the effects of removal of these sources. 
We note a few things to build confidence that our results are not significantly 
affected by the errors due to background subtraction:  
\begin{enumerate}
\item We find that the background sources are all unresolved sources,
which greatly simplifies the deconvolution problem. 
\item The \textit{rms} in the UVSUB maps including regions 
from where sources have been subtracted
are similar to the \textit{rms} in the final map of the Venus. This implies
that any residual flux left behind after cleaning is small enough to not
give rise to any discernible artifacts even at the lowest frequency. 
\item There are few background sources close to Venus. Hence, any contamination 
from them can only be due to the side lobes of the point-spread-function (PSF).
\item When using CASA to make the final maps for the Venus, we cut out the 
appropriate parts of the 1 minute maps and align them to ensure that flux from 
Venus falls in the same pixels. 
This has the consequence that any small residual flux from the background sources will
get smeared in the Venus maps, further reducing any contamination from them. Similarly, 
when using AIPS, though the implementation details differ, the methodology followed ensures 
the residual flux from the background sources will get smeared over a region spanning
the track of the Venus in the sky plane.
\end{enumerate}

\section{RESULTS}
\begin{figure}[H]
 \begin{center}
  \includegraphics[width=0.5\textwidth, center]{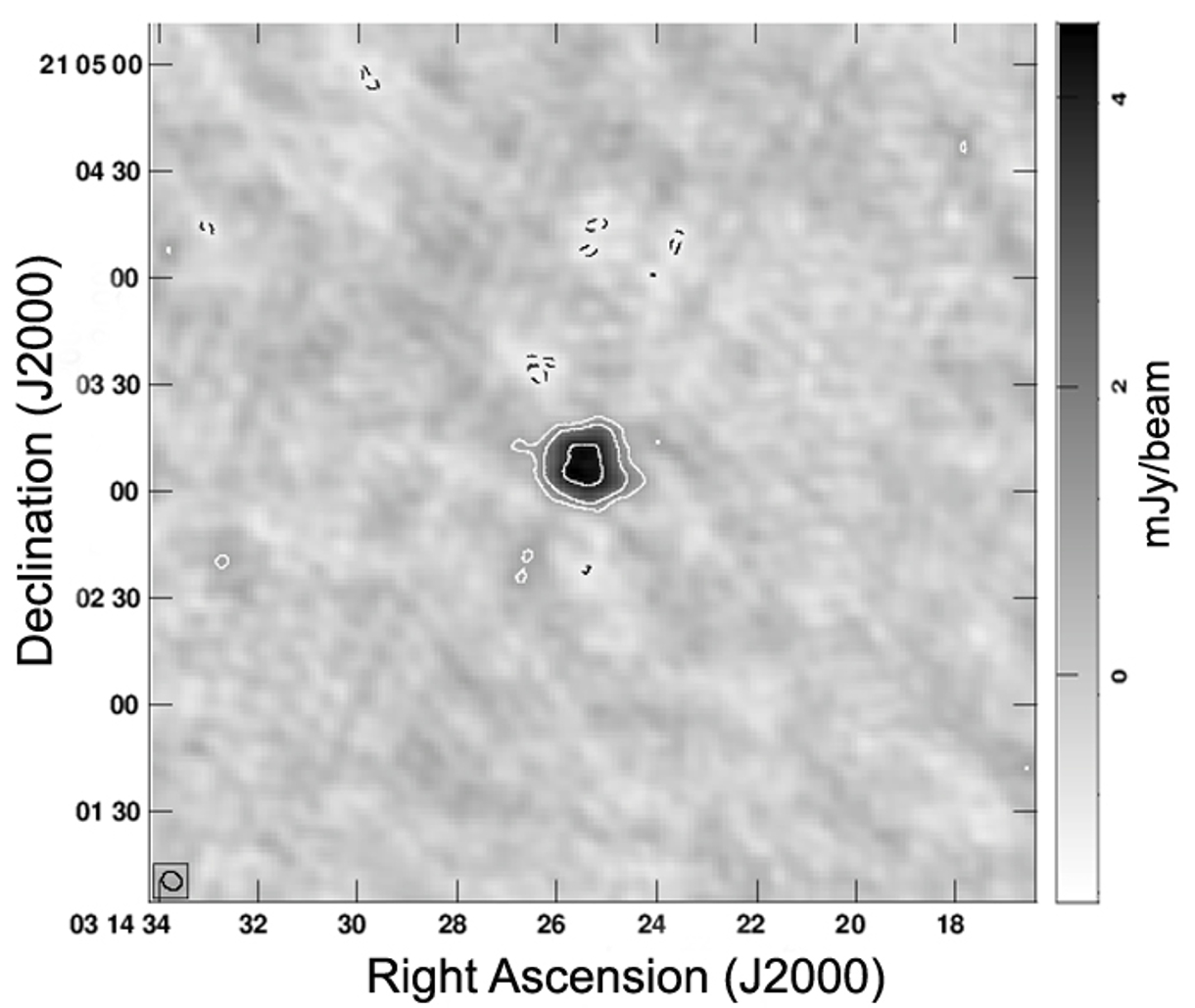}
  \caption{ \label{fig.image_tb606}
	    Image of Venus made at 606.1\,MHz on 26 March 2004 using the GMRT. 
	    Contour values are -2, -1, 1, 2, 4, 6 mJy. Peak flux value is 5.93 
	    mJy and the total flux density is 58.0\,mJy. The \textit{rms} value 
	    of the image = 0.45\,mJy.}
 \end{center}

\end{figure}

\begin{figure}[H]
 \begin{center}
  \includegraphics[width=0.5\textwidth, center]{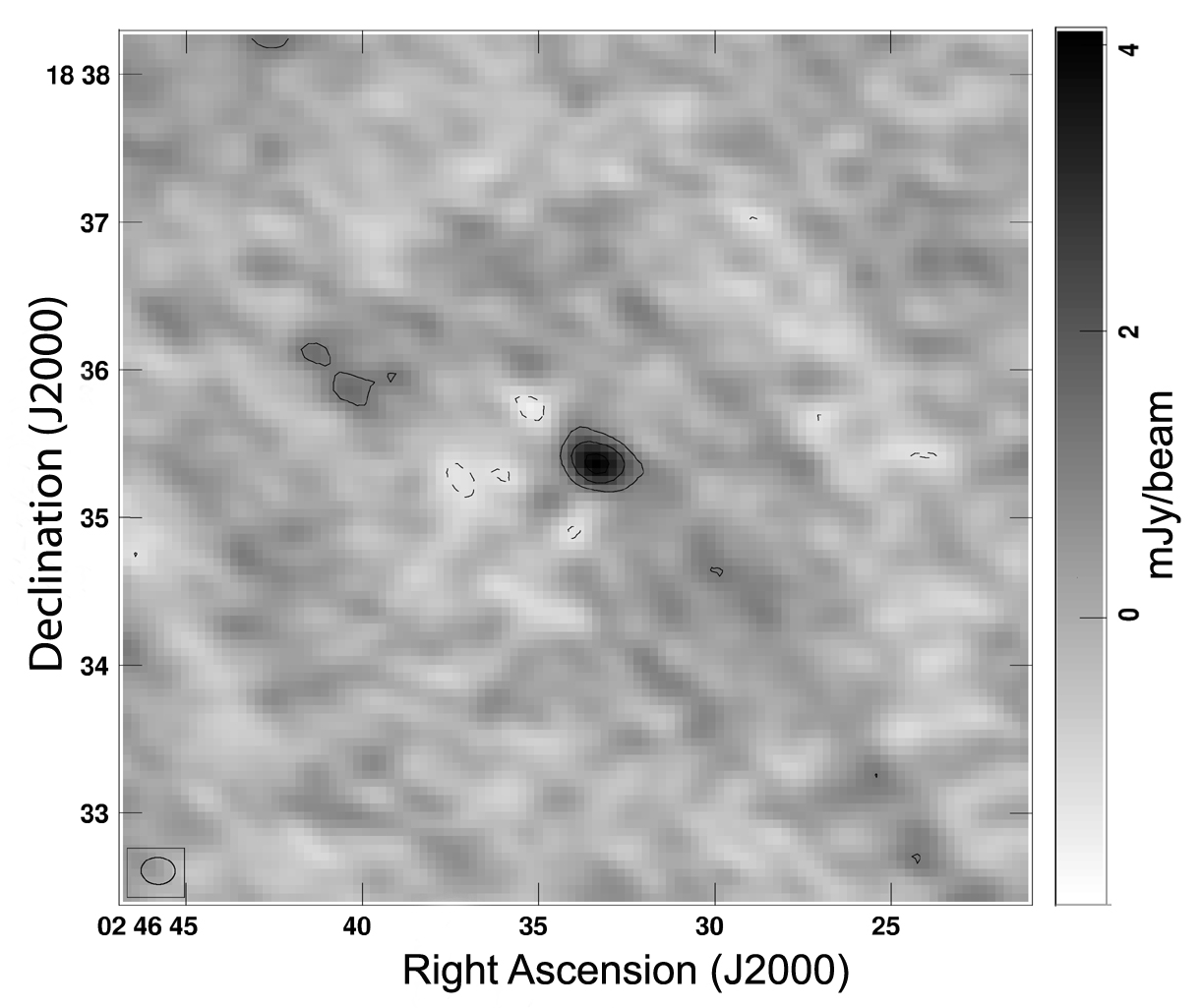}
  \caption{\label{fig.image_tb333}
	   Image of Venus made at 332.9\,MHz on 19 March 2004 using GMRT. 
	   Contour values are -2,-1,2,3,4,6 mJy. 
	   Peak flux value is 4.07\,mJy and the total flux density is 
	   10.6\,mJy. The \textit{rms} value of the image = 0.4\,mJy.}
 \end{center}

\end{figure}

Figure\,\ref{fig.image_tb606} shows the map of Venus at 606.1\,MHz from the observations
of March 26, 2004 and Figure\,\ref{fig.image_tb333} shows
the map of Venus at 332.9\,MHz from the observations of 19 March 2004.
The error on the measured flux density of Venus was obtained by multiplying the 
measured \textit{rms} noise in the background image by $\sqrt{N}$, where $N$ is 
the total surface area of Venus measured in units of the synthesized beam.
We then multiply by a factor of 1.07 to take care of the random and 
systematic errors as discussed earlier.
From the known solid angle subtended by Venus during these observations 
(Table 2), its brightness temperature, $\mathit{T}_\textup{b}$, can be estimated
from the observed values of its flux density using Rayleigh-Jeans law. 

However, an additive correction needs to be applied to the $\mathit{T}_\textup{b}$ thus determined.
To understand its origin, we note that the {\em uv} coverage of any interferometer has a central hole, 
reflecting the absence of baselines shorter than some minimum length.
This leads to the common situation that the peak of the point-spread-function 
(PSF) is surrounded by a shallow negative bowl, or equivalently the interferometer
is not sensitive to brightness distribution at large angular scales.
A practical consequence of this is that the interferometer resolves out the smooth 
Galactic background and when the PSF is convolved with an extended source like Venus, 
the source is observed to be sitting in a bowl of negative flux \citep{Taylor1999}.
Also, the Galactic background radiation gets fully absorbed by Venus \citep{Condon1973}.
Together, they lead to an underestimate in the true value of the $\mathit{T}_\textup{b}$
of Venus by an amount equal to the temperature of the Galactic background, 
$\mathit{T}_\textup{gal}$, which is resolved out by the interferometer and needs
to be added to get the true brightness temperature for Venus, $\mathit{T}_\textup{b,cor}$.
We used the $\mathit{T}_\textup{gal}$ values as discussed in Sec. 3.3.

The $\mathit{T}_\textup{b,cor}$ values for the three GMRT frequencies 
are provided in Table 3. Column\,1 lists the frequency of observation,
column\,2 the date of observation and columns\,3 and 4 are the size and the 
position angle of the synthesized beam, 
respectively. Column\,5 gives the measured \textit{rms} in the map of Venus,
Columns\,6 and 7 are measured flux density and 
the estimated \textit{rms} error of the flux density in the map, respectively.
Column\,8 gives the $\mathit{T}_\textup{b}$ computed from the 
measured flux density and the size of the known sources, the magnitude of the correction
for $\mathit{T}_\textup{gal}$ is given in column\,9. 
Column\,10 gives the final computed values of the brightness temperature of 
Venus, $\mathit{T}_\textup{b,cor}$. In column\,11, the
numbers in bold give the average value of $\mathit{T}_\textup{b}$ for a given frequency.
As discussed in Sec. 3.2, $\sim$7\% error is assumed to account for random and 
other systematic errors.
The lowest frequency data (239.9 MHz) was analyzed using both CASA and AIPS 
which gave similar \textit{rms} in the image plane.
In both the cases, Venus could not be detected in the image.
A 3\,$\sigma$ value is used to place an upper limit of the brightness temperature of Venus
at this frequency and is indicated by $\downarrow$ in Table 3.


\begin{table*}[!t]
  
  \begin{center}
  \caption{Summary of $\mathit{T}_\textup{b}$ measurements (2004)} 
  \centering\small
\begin{tabular}{ccccccccccc}

\\
\hline
  Frequency	& Date     & Beam Size 		& Beam 	   		& Map		& Flux              & $\Delta$ Flux &  $\mathit{T}_\textup{b}$(K)		 & $\mathit{T}_\textup{gal}$(K)	 & $\mathit{T}_\textup{b,cor}$ (K)   & mean 				\\
  (MHz)		&	   &  (arcsec)		& Position 		&\textit{rms}	& (mJy)		    & (mJy)	    &			 & 			 &								     & $\mathit{T}_\textup{b,cor}$	\\	
		&	   &	   		& Angle\,(\textdegree)  &(mJy)		&		    &	 	    &			 &			 &	  	 	     					     &(K)				\\
  \hline
  606.1		& March 26 & 6.5$^{\prime \prime}$ $\times$ 5.1$^{\prime \prime}$ 	& 53.7 		    &0.45		& 58.0 		    & 1.68	    & 540		 & 10$\pm$1.5		 & 550			     & 					\\
  606.0		& March 27 & 5.5$^{\prime \prime}$ $\times$ 4.2$^{\prime \prime}$	& 57.0 		    &0.56		& 54.0 		    & 2.54	    & 492		 & 10$\pm$1.5		 & 502			     & \textbf{526\,$\pm$\,22}		\\
  332.9		& March 19 & 13.8$^{\prime \prime}$ $\times$ 11.0$^{\prime \prime}$	& 87.2		    &0.40		& 10.6		    & 0.65	    & 376		 & 52$\pm$7		 & 428			     &					\\
  332.9 	& March 20 & 13.5$^{\prime \prime}$ $\times$ 11.4$^{\prime \prime}$	& 81.6		    &0.50		& 9.7		    & 0.82	    & 337		 & 52$\pm$7		 & 389			     & \textbf{409\,$\pm$\,33}		\\
  239.9		& March 26 & 15.4$^{\prime \prime}$ $\times$ 13.3$^{\prime \prime}$	& 89.0		    &1.50		& \textdownarrow4.74 & 1.50	    & \textdownarrow282	 & 122$\pm$22		 &\textbf{\textdownarrow426} &					\\
  \hline

\end{tabular}

\end{center}
\end{table*}

\begin{figure}[H]

  \begin{center}
  
  \includegraphics[width=0.55\textwidth ,center]{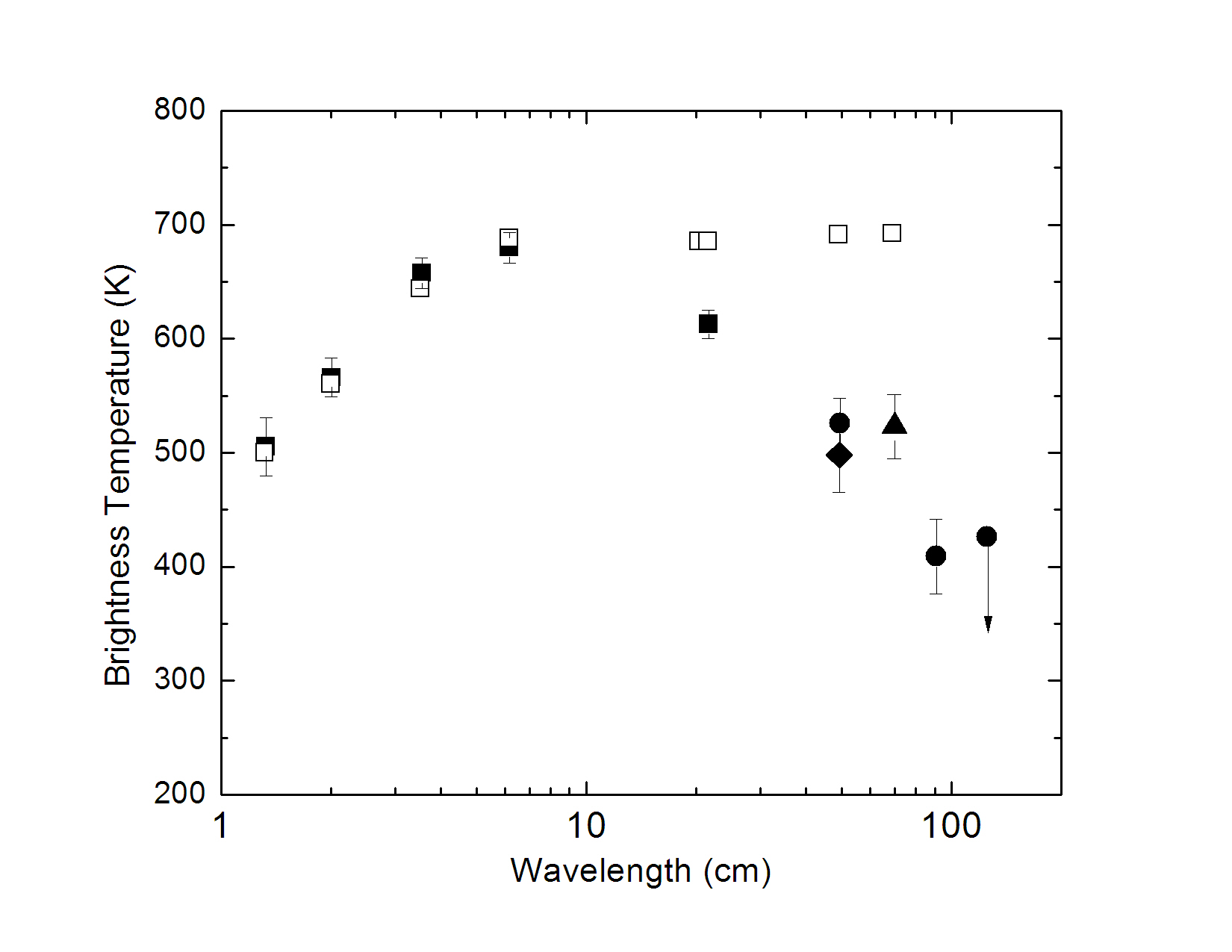}
  \caption{\label{fig.image_tbplot}
	  Plot of derived brightness temperature, $\mathit{T}_\textup{b}$, versus wavelength. 
	  ($\blacksquare$) shows measured values and ($\Box$) shows the model
	  calculations by \citet{Butler2001} using the VLA. The $\mathit{T}_\textup{b}$ value as 
	  obtained by \citet{Muhleman1973} is shown in ($\Diamondblack$),
	  that by \citet{Condon1973} is marked by ($\blacktriangle$) and 
	  ($\medbullet$) shows values obtained using the GMRT (this work). Vertical lines 
	  show $\pm$ one \textit{rms} values.}
	   
 \end{center}

\end{figure}

Figure \ref{fig.image_tbplot} and Table 4 compile all available measurements of brightness temperature 
of Venus in the wavelength range from 0.013 m (22.46 GHz) to 1.25 m (239.3 MHz), including the ones obtained by us.
Figure \ref{fig.image_tbplot} also includes the model by \citet{Butler2001}.
Our observations clearly indicate that $\mathit{T}_\textup{b,cor}$ decreases with
increasing wavelength beyond $\sim$ 0.5 m, in contrast to the model which remains practically flat beyond $\sim$0.06 m.

\begin{table*}[!t]
\label{lasttable}
\begin{center}
  \caption{Derived values of $\mathit{T}_\textup{b}$ as a function of frequency and wavelength}
  \centering\small
  \begin{tabular}{ccccc}
  \\
  
  \hline
  Frequency   	  & Wavelength (m)	&	$\mathit{T}_\textup{b}$ (K)	&	Investigators			& Radio Telescopes		\\
  (MHz)		  &			&					&					&				\\
  \hline	

  
22460    	  &	0.013		&	505.2\,$\pm$\,25.3		&	\citet{Butler2001}		&	VLA			\\
14940   	  &	0.020		&	565.9\,$\pm$\,17.0		&	\citet{Butler2001}		&	VLA			\\
8440    	  &	0.036		&	657.5\,$\pm$\,13.2		&	\citet{Butler2001}		&	VLA			\\
4860   		  &	0.062		&	679.9\,$\pm$\,13.6		&	\citet{Butler2001}		&	VLA			\\
1465		  &	0.204		&	619.0\,$\pm$\,12.4		&	\citet{Butler2001}		&	VLA			\\
1385   		  &	0.217		&	612.8\,$\pm$\,12.3		&	\citet{Butler2001}		&	VLA			\\
606    		  &	0.495		&	526\,$\pm$\,22			&	This work			&	GMRT			\\
608     	  &	0.493		&	498\,$\pm$\,33			&	\citet{Muhleman1973}		&	Owen's valley		\\
430    		  &	0.698		&	523\,$\pm$\,28			&	\citet{Condon1973}		&	Arecibo 300m dish	\\
332.9		  &	0.909		&	409\,$\pm$\,33			&	This work			&	GMRT			\\
239.9		  &	1.250		&	$<$\,426			&	This work			&	GMRT			\\
 \hline	
  \end{tabular}
  \end{center}
\end{table*}


\section{DISCUSSION}

As can be seen from Table 4 and Figure\,\ref{fig.image_tbplot} that the $\mathit{T}_\textup{b}$ 
of Venus obtained from GMRT
observations at longer wavelengths and those reported values by earlier
investigators \citep{Condon1973,Muhleman1973}, are appreciably lower than those observed by 
\citet{Butler2001} at\,cm wavelengths (at frequencies $>$\,1.385\,GHz). 
The detailed model $\mathit{T}_\textup{b}$ values as seen in Figure\,\ref{fig.image_tbplot} 
agree very well with the observed
values by \citet{Butler2001} at shorter wavelengths ($<$\,6\,cm) increasing 
with the same in a log-linear manner with a slope of $\sim$\,40\,K/cm which peaks around 6\,cm.
The observations beyond 11\,cm wavelength again show a decreasing $\mathit{T}_\textup{b}$ values 
in a log-linear form. But the model by \citet{Butler2001} 
predicts the values of $\mathit{T}_\textup{b}$ at lower frequencies (or longer 
wavelengths) to be same as that at $\sim$\,6\,cm. 

\par Very low atmospheric opacity, nearly 0.006 $\pm$ 0.005 at 608\,MHz and about 0.02 
at 1.4\,GHz were reported by \citet{Muhleman1973} and \citet{Butler2001}, 
respectively. As the GMRT observations are at and around these frequencies, it 
can be safely  assumed that the atmosphere is almost transparent at the GMRT frequencies. 
Another possible mechanism responsible 
for the reduction in $\mathit{T}_\textup{b}$ at radiowaves was the presence of selective 
wavelength absorbing ionosphere as suggested by \citet{Kuzmin1964,Kuzmin1967}.
However, this suggestion was ruled out by \citet{Warnock1972} based on the 
Mariner V electron density profile measurements \citep{Herman1971} which showed
a low peak electron density of 5.2 $\times$ 10$^{5}$\,cm$^{-3}$ at an altitude
of 135-140\,km, which was not enough to act as an absorber. 
\citet{Muhleman1973} and \citet{Condon1973} in their independent 
investigations reported the insignificance of the Venusian ionosphere at lower 
frequencies at $\sim$\,608\,MHz and 430\,MHz (70\,cm) so that 
the Venusian atmosphere including the ionosphere can be neglected in 
determining the $\mathit{T}_\textup{b}$ at low frequencies. 
\par The other possible reasons for the reduction in $\mathit{T}_\textup{b}$ with wavelength 
observed in the radio observations at decimeter wavelength could be the
variations in the dielectric constant or temperature.
The radar signal can be significantly affected by the reflection and absorption 
depending on the dielectric properties of the surface medium, whereas the 
scattering is controlled by the surface roughness. 
It must be noted that the temperature plays only a minor role in the 
variation of the radar signal. Based on radar observations at 50\,MHz and 38\,MHz  
(\citeauthor{Klemperer1964},\,\citeyear{Klemperer1964}; 
\citeauthor{James1964},\,\citeyear{James1964}; 
\citeauthor{James1967},\,\citeyear{James1967}),
\citet{Condon1973} have ruled out a drastic variation in the values of the dielectric constant
of the Venusian regolith at least up to several tens of meters. The dielectric constant 
measurements of typical planetary rocks including basalts at 450\,MHz and 
35\,GHz by \citet{Campbell1969} revealed no significant variation in the 
dielectric properties with frequency. They have also ascertained the absence of
absorbing lines which can alter the dielectric values in between these two 
frequencies.
\par \citet{Warnock1972} using their two-layer subsurface model tried to explain the 
reduced radar reflectivity at centimeter wavelength and reduced brightness 
temperature at decimeter wavelength. The best fit in explaining the reduced 
reflectivity at cm wavelength and $\mathit{T}_\textup{b}$ at decimeter wavelength was obtained when a 
two-layer model consisting of a layer with $\varepsilon$ = 1.5 overlaying another
layer of $\varepsilon$ = 8.31 was chosen. However, a better fit to the 
observation
for $\lambda$ $>$ 15\,cm was obtained when a possible decreasing radiating temperature 
with increasing subsurface depth was assumed. This is expected as at longer wavelengths
the emission is dominated by the deeper subsurface layers owing to deeper
penetration at these frequencies. They did not further probe for a satisfactory
explanation for the reduction in planetary regolith temperature with depth. 
When emissivity of an object is close to unity at a 
particular wavelength, its $\mathit{T}_\textup{b}$ approaches to its physical temperature.
In the case of Venus, the optical depth of its dense atmosphere at decimeter
and meter wavelengths is much lower than unity and the 
observed emission is expected to be generated from its surface and subsurface 
with certain depth. 
\par The reduction in $\mathit{T}_\textup{b}$ due to
the atmosphere, ionosphere and the variation of dielectric values with decreasing frequencies 
are not expected to be significant. 
Radiative transfer model is an effective tool for computing thermal emission
at microwave and radio wavelengths by accounting for the detailed variation of 
temperature and dielectric properties with the depth of the terrain as well as
with altitude of the atmosphere of Venus.
Further studies are needed to explain the lower values of $\mathit{T}_\textup{b}$ at frequencies 
$<$\,1\,GHz (meter wavelengths), where emission arises 
predominantly from a region further down the surface owing to deeper penetration.

\section{CONCLUSION}

The first interferometric imaging observations of Venus at frequencies below
620\,MHz are presented here.
These observations of thermal emission from Venus were conducted using the GMRT.
The analyses of these data revealed that brightness temperature of Venus 
decreases with increasing wavelength 
as 526\,K\,$\pm$\,22, 409\,K\,$\pm$\,33,  and $<$\,426\,K at 606, 332.9, and 
239.9\,MHz, respectively. These values are consistent with values 
of about 498\,K and 523\,K measured at 608 and 430\,MHz, respectively by previous workers 
during the 1970s, but are much lower than those measured at
higher frequencies, \textit{e.g.}, 679.9\,K\,$\pm$\,13.6 at 4.86\,GHz using the 
VLA. The microwave observations (cm wavelengths) of 
$\mathit{T}_\textup{b}$ of Venus has been explained earlier by considering emission from its 
atmosphere and surface. The observed variation of the $\mathit{T}_\textup{b}$ at low microwave 
frequencies ($<$\,1\,GHz) can only be explained with further radiative transfer
studies, as in this frequency
regime, the emission is dominated by the surface/subsurface of the planetary regolith.

\label{lastpage}
\section{ACKNOWLEDGEMENT}
Authors thank Dr. Dharam Vir Lal, NCRA-TIFR, Dr.\,K Krishnamoorthy, former Director, 
SPL, VSSC and Dr. Nizy Mathew,  SPL, VSSC for
many valuable discussions. We thank the staff of the GMRT who made these 
observations possible.
The GMRT is run by the National Centre for Radio Astrophysics of the Tata 
Institute of Fundamental Research. Finally, authors thank the anonymous referees
for their constructive comments and valuable suggestions.
Mr. Nithin Mohan is supported by ISRO Research Fellowship.


\section{REFERENCE}

\begin{thebibliography}{41}
\expandafter\ifx\csname natexlab\endcsname\relax\def\natexlab#1{#1}\fi
\expandafter\ifx\csname url\endcsname\relax
  \def\url#1{\texttt{#1}}\fi
\expandafter\ifx\csname urlprefix\endcsname\relax\def\urlprefix{URL }\fi

\bibitem[{{Baars} et~al.(1977){Baars}, {Genzel}, {Pauliny-Toth}, and
  {Witzel}}]{Baars1977}
{Baars}, J. W.~M., {Genzel}, R., {Pauliny-Toth}, I. I.~K., {Witzel}, A., 1977.
  The {A}bsolute {S}pectrum of {C}as {A}. {A}n {A}ccurate {F}lux {D}ensity
  {S}cale and a {S}et of {S}econdary {C}alibrators. Astron. Astrophys.
  61~(99-106).

\bibitem[{{Basilevsky} et~al.(1986){Basilevsky}, {Pronin}, {Ronca},
  {Kryuchkov}, {Sukhanov}, and {Markov}}]{Basilevsky1986}
{Basilevsky}, A.~T., {Pronin}, A.~A., {Ronca}, B., {Kryuchkov}, V.~P.,
  {Sukhanov}, A.~L., {Markov}, S., 1986. {Styles of {T}ectonic {D}eformations
  on Venus: {A}nalysis of {V}enera 15 and 16 {D}ata}. J. Geophys. Res.
  91~(399-411).

\bibitem[{{Butler} et~al.(2001){Butler}, {Steffes}, {Suleiman}, {Koldoner}, and
  {Jenkins}}]{Butler2001}
{Butler}, B.~J., {Steffes}, P.~G., {Suleiman}, S.~H., {Koldoner}, M.~A.,
  {Jenkins}, J.~M., 2001. {Accurate and {C}onsistent {M}icrowave {O}bservations
  and their {I}mplications}. Icarus 154~(226-238).

\bibitem[{{Campbell}(1994)}]{Campbell1994}
{Campbell}, B.~A., 1994. {Merging {E}missivity and {SAR} {D}ata for {A}nalysis
  of {V}enus {D}ielectric {P}roperties}. Icarus 112~(187-203).

\bibitem[{{Campbell} et~al.(1989){Campbell}, {Head}, {Hine}, {Harmon},
  {Senske}, and {Fisher}}]{Campbell1989}
{Campbell}, D.~N., {Head}, J.~W., {Hine}, A.~A., {Harmon}, J.~K., {Senske},
  D.~A., {Fisher}, P.~C., 1989. {Styles of {V}olcanism on {V}enus: {N}ew
  {A}recibo {H}igh {R}esolution {R}adar {D}ata}. Science 246~(373-377).

\bibitem[{{Campbell} and {Ulrichs}(1969)}]{Campbell1969}
{Campbell}, M.~J., {Ulrichs}, J., 1969. {Electrical {P}roperties of {R}ocks and
  {T}heir {S}ignificance for {L}unar {R}adar {O}bservations}. J. Geophys. Res.
  74~(5867-5881).

\bibitem[{{Carpenter}(1964)}]{Carpenter1964}
{Carpenter}, R.~L., 1964. {Study of {V}enus {S}urface by {C}.{W}. {R}adar}.
  Astron. J. 69~(1-11).

\bibitem[{{Chengalur}(2013)}]{Chengalur2013}
{Chengalur}, J.~N., 2013. {NCRA} {T}echnical {R}eport, {NCRA}/{COM}/{OD},.
  Tech. rep., {N}ational {C}entre for {R}adio {A}strophysics, {P}une 411007,
  {I}ndia.

\bibitem[{{Condon} et~al.(1973){Condon}, {Jauncey}, and {Yerbury}}]{Condon1973}
{Condon}, J.~J., {Jauncey}, D.~L., {Yerbury}, M.~J., 1973. {The {B}rightness
  {T}emperature of {V}enus at 70 {C}entimeters}. Astrophys. J. 183~(1075-1080).

\bibitem[{{Fjeldbo} et~al.(1971){Fjeldbo}, {Kliore}, and
  {Eshleman}}]{Fjeldbo1971}
{Fjeldbo}, G., {Kliore}, A.~J., {Eshleman}, V.~R., 1971. {The {N}eutral
  {A}tmosphere of {V}enus as {S}tudied {W}ith the {M}ariner {V} {R}adio
  {O}ccultation {E}xperiment}. Astron. J. 73~(123-140).

\bibitem[{{Florenskii} et~al.(1982){Florenskii}, {Bazilevsky}, {Kruchkyov},
  {Kuzmin}, {Nikolaeva}, {Pronin}, {Selivanov}, {Naraeva}, and
  {Tyuflin}}]{Florenskii1982}
{Florenskii}, K.~P., {Bazilevsky}, A.~T., {Kruchkyov}, V.~P., {Kuzmin}, O.~V.,
  {Nikolaeva}, {Pronin}, A A~{Chernaya}, I.~M., {Selivanov}, A.~S., {Naraeva},
  M.~K., {Tyuflin}, Y.~S., 1982. {Analysis of the {P}anoramas of the {V}enera
  13 and {V}enera 14 {L}anding {S}ites}. Sov. Astron. Lett. 8~(233-234).

\bibitem[{{Ford} and {Pettengill}(1983)}]{Ford1983}
{Ford}, P.~G., {Pettengill}, G.~H., 1983. {Venus: {G}lobal {S}urface {R}adio
  {E}missivity}. Science 220~(1379-1381).

\bibitem[{{Goldstein} and {Carpenter}(1963)}]{Goldstein1963}
{Goldstein}, R.~M., {Carpenter}, R.~L., 1963. {Rotation of {V}enus: {P}eriod
  {E}stimated from {R}adar {M}easurements}. Science 139~(910-911).

\bibitem[{{Haslam} et~al.(1983){Haslam}, {Salter}, {Stoffel}, and
  {Wilson}}]{Haslam1982}
{Haslam}, C. G.~T., {Salter}, C.~J., {Stoffel}, H., {Wilson}, W.~E., 1983. {A
  408 {M}{H}z {A}ll-{S}ky {C}ontinuum {S}urvey {I}{I} - The {A}tlas of
  {C}ontour {M}aps}. Astron. Astrophys. Suppl. Series 47~(1-142).

\bibitem[{{Herman} et~al.(1971){Herman}, {Hartle}, and {Bauer}}]{Herman1971}
{Herman}, J., {Hartle}, R., {Bauer}, S., 1971. The {D}ayside {I}onosphere of
  {V}enus. The Planet. Space Sci. 19.

\bibitem[{{James} and {Ingalls}(1967)}]{James1964}
{James}, J., {Ingalls}, R., 1967. Radar {O}bservation of {V}enus at 38
  {M}c/sec. The Astron. Journ. 72.

\bibitem[{{James} et~al.(1967){James}, {Ingalls}, and {Rainville}}]{James1967}
{James}, J., {Ingalls}, R., {Rainville}, L., 1967. Radar {E}chos from {V}enus
  at 38 {M}c/sec. The Astron. Journ. 72.

\bibitem[{{Klemperer} and {Bowles}(1964)}]{Klemperer1964}
{Klemperer}, W.K.and~{Ochs}, G., {Bowles}, K., 1964. Radar {E}chos from {V}enus
  at 50 {M}c/sec. The Astron. Journ. 69.

\bibitem[{Kuzmin(1964)}]{Kuzmin1964}
Kuzmin, A., 1964. Radio {P}hysical {I}nvestigations of {V}enus. In: Physics,
  {A}ll - {U}nion {I}nstitute of {S}cientific and {T}echnical {I}nformation.
  Academy of {S}cience {USSR}, {M}oscow.

\bibitem[{Kuzmin(1967)}]{Kuzmin1967}
Kuzmin, A., 1967. Concerning a {M}odel of {V}enus with {C}old {A}bsorbing
  {A}tmosphere. Izv. Vyssh. Ueheb. Zaved. Radiofiz. 7, 1021--1031.

\bibitem[{{Kuzmin}(1983)}]{Kuzmin1983}
{Kuzmin}, A.~D., 1983. Radio {A}stronomical {S}tudies of {V}enus. In: Venus.
  University of Arizona Press, Tucson, Arizona, pp. 37--44.

\bibitem[{{Markiewicz} et~al.(2007){Markiewicz}, {Titov}, {Limaye}, {Keller},
  {Ignatiev}, {Jaumann}, {Thomas}, {Michalik}, {Moissl}, and
  {Russo}}]{Markiewicz2007}
{Markiewicz}, W.~J., {Titov}, D.~V., {Limaye}, S.~S., {Keller}, H.~U.,
  {Ignatiev}, N., {Jaumann}, R., {Thomas}, N., {Michalik}, H., {Moissl}, R.,
  {Russo}, P., 2007. {Morphology and {D}ynamics of the {U}pper {C}loud {L}ayer
  of {V}enus.} Nature 450~(633-636).

\bibitem[{{Marov}(1978)}]{Marov1978}
{Marov}, M.~Y., 1978. {Results of {V}enus {M}issions}. Annu. Rev. Astron.
  Astrophys. 16~(141-169).

\bibitem[{{Muhleman} et~al.(1973){Muhleman}, {Berge}, and
  {Orton}}]{Muhleman1973}
{Muhleman}, D.~O., {Berge}, G.~L., {Orton}, G.~S., 1973. {The {B}rightness
  {T}emperature of {V}enus and the {A}bsolute {F}lux-{D}ensity {S}cale at 608
  {M}{H}z.} Astrophys. J. 183~(1081-1085).

\bibitem[{{Muhleman} et~al.(1979){Muhleman}, {Orton}, and
  {Berge}}]{Muhleman1979}
{Muhleman}, D.~O., {Orton}, G.~S., {Berge}, G.~L., 1979. {A {M}odel of the
  {V}enus {A}tmosphere from {R}adio, {R}adar, and {O}ccultation
  {O}bservations}. Astrophys. J. 234~(733-745).

\bibitem[{{Perley} and {Butler}(2013)}]{Perley2013}
{Perley}, R.~A., {Butler}, B.~J., 2013. An {A}ccurate {F}lux {D}ensity {S}cale
  {F}rom 1 to 50 {G}{H}z. Astrophys. J. Supp. Series, 204:19(20 pp).

\bibitem[{{Pettengill} et~al.(1980){Pettengill}, {Eliason}, {Ford}, {Loriot},
  {Masursky}, and {McGill}}]{Pettengill1980}
{Pettengill}, G.~H., {Eliason}, E., {Ford}, P.~G., {Loriot}, G.~B., {Masursky},
  H., {McGill}, G.~E., 1980. {Pioneer {V}enus {R}adar {R}esults: {A}ltimetry
  and {S}urface {P}roperties}. J. Geophys. Res. 85~(8261-8270).

\bibitem[{{Pettengill} et~al.(1988){Pettengill}, {Ford}, and
  {Chapman}}]{Pettengill1988}
{Pettengill}, G.~H., {Ford}, P.~G., {Chapman}, B.~D., 1988. {Venus: {S}urface
  {E}lectromagnetic {P}roperties}. J. Geophys. Res. 93~(14,881-14,892).

\bibitem[{{Pettengill} et~al.(1991){Pettengill}, {Ford}, {Johnson}, {Raney},
  and {Soderblom}}]{Pettengill1991}
{Pettengill}, G.~H., {Ford}, P.~G., {Johnson}, W. T.~K., {Raney}, R.~K.,
  {Soderblom}, L.~A., 1991. {Magellan: {R}adar {P}erformance and {D}ata
  {P}roducts}. Science 252~(260-265).

\bibitem[{{Pettengill} et~al.(1992){Pettengill}, {Ford}, and
  Wilt}]{Pettengill1992}
{Pettengill}, G.~H., {Ford}, P.~G., Wilt, R.~J., 1992. {Venus {R}urface
  {R}adiothermal {E}mission as {O}bserved by {M}agellan}. J. Geophys. Res.
  97~(13,091-13,102).

\bibitem[{{Phillips} and {Malin}(1983)}]{Phillips1983}
{Phillips}, R.~J., {Malin}, M.~C., 1983. The interior of venus and tectonic
  implications. In: Venus. University of Arizona Press, Tucson, Arizona, pp.
  159--214.

\bibitem[{{Prasad} and {Chengalur}(2012)}]{Prasad2012}
{Prasad}, J., {Chengalur}, J.~N., 2012. F{LAGCAL}: a {F}lagging and
  {C}alibration {P}ackage for {R}adio {I}terferometric {D}ata. Exp. Astron.
  33~(157-171).

\bibitem[{{Reich} and {Reich}(1988)}]{Reich1988}
{Reich}, P., {Reich}, W., 1988. Spectral index variations of the galactic radio
  continuum emission - evidence for a galactic wind. Astron. {A}strophys. 196,
  211--226.

\bibitem[{{Scaife} and {Heald}(2012)}]{Scaife2012}
{Scaife}, A. M.~M., {Heald}, G.~H., 2012. A {B}roadband {F}lux {S}cale for
  {L}ow {F}requency {R}adio {T}elescopes. Mon. Not. R. Astron. Soc.
  423~(30-34).

\bibitem[{{Seiff} et~al.(1980){Seiff}, {Kirk}, {Young}, {Blanchard}, {Findlay},
  {Kelley}, and {Soreruer}}]{Seiff1980}
{Seiff}, A., {Kirk}, D.~B., {Young}, R.~E., {Blanchard}, R.~C., {Findlay},
  J.~T., {Kelley}, G.~M., {Soreruer}, S.~C., 1980. {Measurements of the
  {T}hermal {S}tructure and {T}hermal {C}ontrasts in the {A}tmosphere of
  {V}enus, and {R}elated {D}ynamical {O}bservations: {R}esults from the four
  {P}ioneer {V}enus probes}. J. Geophys. Res. 85~(7903-7933).

\bibitem[{{Sinclair} et~al.(1970){Sinclair}, {Basart}, {Buhl}, {Gale}, and
  {Liwshitz}}]{Sinclair1970}
{Sinclair}, A. C.~E., {Basart}, J.~P., {Buhl}, D., {Gale}, W.~A., {Liwshitz},
  M., 1970. {Preliminary {R}esults of {I}nterferometric {O}bservations of
  {V}enus at 11.1 cm {W}avelength}. Radio Science 5~(347-354).

\bibitem[{{Surkov}(1983)}]{Surkov1983}
{Surkov}, Y.~A., 1983. Studies of venus rocks by veneras 8,9 and 10. In: Venus.
  University of Arizona Press, Tucson, Arizona, pp. 154--158.

\bibitem[{{Swarup} et~al.(1991){Swarup}, {Ananthkrishnan}, {Kapahi}, {Rao},
  {Subrahmanya}, and {Kulkarni}}]{Swarup1991}
{Swarup}, G., {Ananthkrishnan}, S., {Kapahi}, V.~K., {Rao}, A.~P.,
  {Subrahmanya}, C.~R., {Kulkarni}, V.~K., 1991. {The {G}iant {M}etrewave
  {R}adio {T}elescope}. Current {S}cience 60~(90-105).

\bibitem[{{Taylor} et~al.(1999){Taylor}, {Carilli}, and {Perley}}]{Taylor1999}
{Taylor}, G.~B., {Carilli}, C.~L., {Perley}, R.~A., 1999. Synthesis Imaging in
  Radio Astronomy II. Vol. 180. Astronomical Society of the Pacific Conference
  Series.

\bibitem[{{Vinogradov} et~al.(1976){Vinogradov}, {Florenskii}, {Bazilevskii},
  and {Selivanov}}]{Vinogradov1976}
{Vinogradov}, A.~P., {Florenskii}, K.~P., {Bazilevskii}, A.~T., {Selivanov},
  A.~S., 1976. {First {P}anoramic {P}ictures of {V}enus: {P}reliminary {I}mage
  {A}nalysis}. Sov. Astron. Lett. 2~(67-71).

\bibitem[{{Warnock} and {Dickel}(1972)}]{Warnock1972}
{Warnock}, W.~W., {Dickel}, J.~R., 1972. {Venus: {M}easurements of {B}rightness
  {T}emperatures in the 7-15-cm {W}avelength {R}ange and {T}heoretical {R}adio
  and {R}adar {S}pectra for a two-layer {S}ubsurface {M}odel}. Icarus
  17~(682-691).

\end{thebibliography}
\bibliographystyle{elsarticle-harv}
 
\end{document}